\documentclass[useAMS,usenatbib]{mn2e}
\usepackage{graphics}
\usepackage{graphicx}
\usepackage{amssymb}
\title[Monte-Carlo simulation of stellar intensity interferometry]{Monte-Carlo simulation of stellar intensity interferometry}
\author[J. Rou, P.~D. Nu{\~n}ez, D. Kieda, and S. LeBohec]{J. Rou$^{1}$\thanks{Email address: janvida.rou@utah.edu }, P.~D. Nu{\~n}ez$^{1}$, D. Kieda$^{1}$, and S. LeBohec$^{1}$\\
$^{1}$Department of Physics \& Astronomy, University of Utah, 115 South 1400 East, Salt lake City, UT 84112-0830, USA}
\begin{document}

\date{Accepted ???? ?? ??. Received ???? ?? ??; in original form ???? ?? ??}

\pagerange{\pageref{firstpage}--\pageref{lastpage}} \pubyear{2012}

\maketitle

\label{firstpage}

\begin{abstract}
Stellar intensity interferometers will achieve stellar imaging with a tenth of a milli-arcsecond resolution in the optical band by taking advantage of the large light collecting area and broad range of inter-telescope distances offered by future gamma-ray Air Cherenkov Telescope (ACT) arrays. Up to now, studies characterizing the capabilities of intensity interferometers using ACTs have not accounted for realistic effects such as telescope mirror extension, detailed photodetector time response, excess noise, and night sky contamination. In this paper, we present the semi-classical quantum optics Monte-Carlo simulation we developed in order to investigate these experimental limitations. In order to validate the simulation algorithm, we compare our first results to models for sensitivity and signal degradation resulting from mirror extension, pulse shape, detector excess noise, and night sky contamination.

\end{abstract}



\section{Introduction and general principles}
\label{intro}
\subsection{Intensity Interferometry}
Stellar Intensity Interferometry (SII) is an experimental method for measuring the angular diameters and acquiring high resolution images of stars. In a stellar intensity interferometer, the light from a star is received by two or more telescopes separated by a baseline that may range from tens of meters to kilometers, allowing for the resolution of surface features ranging from less than 0.1 milli-arcsecond (mas) (longest baseline) to 10 mas (shortest baseline) at visible wavelengths. Intensity interferometry relies on the correlation between the light intensity fluctuations recorded by different telescopes \citep{brown1957}. The fluctuations contain two components: the shot noise and the wave noise. The dominant component is the shot noise, which is the random fluctuation associated with the photon statistics, and which is uncorrelated between telescopes. The smaller component is the wave noise, which can be interpreted as the beating between different Fourier components of the light reaching the different telescopes. The wave noise shows correlation between different telescopes provided there is some degree of mutual coherence in the light. 

The light intensity correlation between receivers $1 \, \& \, 2$ is measured as the time-integrated product of the fluctuations $\delta i_1\ \&\ \delta i_2$ in the photodetector currents. Although higher order correlations can be of interest (\cite{fontana}; \cite{gamo}; \cite{sato}; \cite{ralston}; \cite{ofir}, \citeyearpar{ribak2}, in this paper we will restrict ourselves to two-point correlations. Still, the Monte-Carlo simulation approach described here could also be directly used for higher order correlation studies.

In the case of a thermal light source and an ideal intensity interferometer, the two-point correlation is equal to the squared degree of coherence $|\gamma|^2$ of the light at the two telescopes:
\begin{equation}
|\gamma|^2=\frac{\langle \delta i_1 \cdot \delta i_2 \rangle}{\langle i_1 \rangle \langle i_2 \rangle}
\end{equation}
where $\langle \cdot\cdot\cdot \rangle$ represents a time average.  According to the van Cittert-Zernike theorem (\cite{cittert}; \cite{zernike1938}), the complex degree of coherence $\gamma$ is the normalized Fourier transform of the source radiance. One difficulty associated with image reconstruction using SII is that the measurable quantity is $|\gamma|^2$, implying that the phase of the Fourier transform is lost. Methods of phase recovery and image reconstruction have recently been developed and investigated (\citet{nunez_aug_2011} and references therein) and Nu{\~n}ez showed that details of stellar surface features can be reconstructed with a relatively high degree of accuracy \citep{nunez_2010}. SII can provide stellar imaging to investigate various topics of interest including stellar rotation, limb darkening \citep{nunez_2012}, mass loss in Be-stars, surface temperature inhomogeneities \citep{dravins}, and binary systems.

It was shown by Robert Hanbury Brown and Richard Twiss that the signal $S$ ($\mathrm{photon^2\,s^{-2}}$) in an intensity interferometry measurement is 
\begin{equation}
\label{eq:sgnl}
S = \langle \delta i_1 \cdot \delta i_2 \rangle = A^2 \alpha^2 \eta^2 |\gamma|^2 \Delta \nu \Delta f
\end{equation}
where $A$ is the effective light collecting area of the telescopes, $\alpha$ is the quantum efficiency of the photodetectors, $\eta$ $\mathrm{(photon \cdot m^{-2} \cdot s^{-1} \cdot Hz^{-1})}$ is the spectral density of the light, $\Delta f$ is the signal bandwidth of the photodetectors and electronics, $\Delta \nu$ is the optical bandwidth of the light, and $T$ is the integration time \citep{hbt_book}. The noise $N$ ($\mathrm{photon^2\,s^{-2}}$) is
\begin{equation}
\label{eq:noise}
N = \sqrt{2} \, A \, \alpha \, \eta \, \Delta \nu \, (\Delta f / T)^{1/2}.
\end{equation}

Therefore the signal-to-noise ratio SNR for an intensity correlation measurement with an ideal system is
\begin{equation}
\label{SNR}
SNR=\frac{S}{N}=A\,\alpha\, \eta \,|\gamma|^2 \sqrt{\Delta\,f\,T/2}
\end{equation}
It is worth noting that the SNR is independent of the optical bandwidth $\Delta \nu$ used for the observations. This expression for the SNR only accounts for the photon statistics, previously referred to as the shot noise. The factor of $\sqrt{2}$ accounts for the fact that starlight is non-polarized. If we consider fully polarized light then the SNR is increased by a factor of $\sqrt{2}$.  

There is increasing interest in SII because of the relative ease to achieve long baselines as well as the advantages it may offer in terms of cost if implemented in conjunction with existing or future very high energy (VHE) gamma-ray ACT arrays (\cite{trevor}; \cite{1204.3624}). ACTs operate by taking advantage of the Cherenkov light flashes produced by atmospheric showers. Because the atmospheric Cherenkov light is very faint, observations of VHE gamma-rays require large light collectors ($A \sim 100 \, \mathrm{m^2}$) and are restricted to nights with little or no moonlight. During moonlit nights, when ACT arrays are less effective for gamma-ray observations, they could be used for SII observations through narrow optical bandwidths. The implication is that SII measurements could be performed with ACT arrays with minimal interference with the VHE observation programs while increasing the scientific output of these instruments.

Arrays such as the Cherenkov Telescope Array (CTA), which could consist of up to $\sim$100 telescopes \citep{cta3} could provide thousands of different baselines simultaneously and could offer detailed imaging capabilities. In previous studies characterizing SII sensitivity and imaging capabilities, it has been assumed that the detectors are point-like in size so the van Cittert-Zernike theorem applies directly and the degree of correlation between the intensity fluctuations is strictly proportional to the squared magnitude of the Fourier transform of the source radiance. However, when the telescope aperture becomes comparable to the baseline separating the telescopes, then the light may no longer be regarded as fully coherent across individual apertures \citep{hbt_book} and the correlation data then departs from a pure Fourier transform. 

Additionally, other instrumentation-related systematic effects have been neglected in previous studies. For example, electronic artifacts such as single photon response pulse profile and excess noise affect the signal or contribute to the degradation of the SNR. Also, the effect of the night sky background light integrated in the point spread function (PSF) of the light collector and the profile of the optical band pass have only been qualitatively or very approximately taken into account \citep{opticalii}. Furthermore, high speed digitization electronic systems are considered to be used for SII applications as they offer the flexibility of offline signal correlation analysis. The development of data analysis algorithms for such systems requires realistic simulated data. Finally, other effects such as mirror non-isochronism and inaccuracies in time delay lines and star tracking can be investigated. All of these aspects are important for the performance characterization, data analysis preparation, and the design and deployment of an SII observatory. 

In this paper we present a Monte-Carlo simulation of a semi-classical quantum description of light and a simple instrumentation model we developed in order to investigate the above mentioned instrumentation-related effects. Section~\ref{simul} describes the simulation approach to achieve the proper photon statistics at the different telescopes and it also includes models of instrumentation effects such as the single photon response pulse and excess noise. Section~\ref{app} presents a few simulation applications characterizing instrumentation related effects, concentrating on the finite telescope diameter, the single photon pulse shape, the excess noise, and the night sky contamination. Since there is no actual data available to test our simulations against, in each case we compare the results to simple models corresponding to ideal cases to obtain further validation of our approach. Finally, the findings are summarized in Section~\ref{conclu}. 

\section{Photon level Monte-Carlo simulation of SII}
\label{simul}
\subsection{Thermal light source and telescope signals}
\label{signals}
We model a stellar light source with a total photon flux $\Phi_\star\ (\mathrm{photon \cdot m^{-2} \cdot s^{-1})}$ within a given optical bandwidth $\Delta \nu$  as a collection of $M$ discrete sources,
each contributing an equal flux  $\Phi_M={\Phi_\star}/{M}$. Each point source is defined by a wave amplitude $\mathcal{A}_j$, an angular frequency $\omega_j$, a phase $\phi_j$, and an angular position $\theta_j=(\theta_x,\theta_y)$. At a given time, the phase $\phi_j$ of the light is taken randomly and uniformly from $[0,2\pi]$ and the amplitude $\mathcal{A}_j$ is taken randomly from a Gauss deviate ~\citep{mandel} of mean $\overline{ \mathcal{A}_j}=0$ and variance $\overline{ \mathcal{A}_j^2}={\Phi_M}/{c}$, where $c$ is the speed of light. In order to make the simulation closer to the continuous distribution of a realistic stellar source, at each time-step we also randomly set the angular frequency $\omega_j$ and angular position $\theta_j=(\theta_x,\theta_y)$ of each point source from distributions corresponding to the spectral density spectrum, radiance, and size of the simulated stellar source.

Each telescope mirror is modeled as a set of small light collecting elements of area $dA_k$ centered on position $X_k=(x_k,y_k)$. Note that ``small'' in this context means that the mutual degree of coherence between any two points within any single area element is maximal. 

The average number of photons emitted by the star and incident on one area element $k$ of the telescope during a time-step $\delta t$ can be written:
\begin{equation}\label{interfequ}
\overline{d\mu_k}=dA_k\cdot \delta t \cdot c \left| \sum_ j \mathcal A_j e^{i\left(\omega_j({{\theta_j\cdot X_k}\over{c}})+\phi_j\right)}\right|^2
\end{equation}
Throughout this paper,  the notation $\overline{\cdot \cdot \cdot}$ denotes a statistical average while $\langle \cdot \cdot \cdot \rangle$ denotes a time average.

The average number of star photons collected by telescope $i$ during a given time-step can then be written: 
\begin{equation}
\bar \mu_i=\sum_k \overline{d\mu_k}.
\end{equation}
Note that $\langle \bar \mu_i \rangle$ is the time average of $\bar \mu_i$ and $\langle \bar \mu_i \rangle=\bar n_{\star_i}$ where $\bar n_{\star_i}=\Phi_\star \, A_i \, \delta t$ is the average number of photons emitted by the star and collected by the telescope in one time-step. The summation of the waves prior to taking the magnitude is responsible for the non-Poisson distributed light intensity fluctuations, previously referred to as wave noise, displaying correlation between the different telescopes.

The actual number of photons reaching telescope $i$ in a time $\delta t$ consists of light from the star as well as stray background light and can be written as
\begin{equation}
n_{T_i}=P(\bar \mu_i + \bar n_{BG_i})=P(\bar \mu_i + \beta \, \bar n_{\star_i})
\end{equation}
where $P(m)$ represents a Poisson deviate of mean and variance $m$ and the night sky contamination is defined in terms of the average source radiance ($n_{BG_i}=\beta \bar n_{\star_i}$ where $\beta$ is a dimensionless factor which sets the amount of stray light).  

In principle, with the description provided thus far, we can simulate any intensity interferometry signal, taking the time-step to be smaller than the coherence time $\tau_c \ (\sim 10^{-5} \, \mathrm{ns})$ of the light. However, in practice, it is desirable to be able to use a time-step $\delta t\gg\tau_c$ comparable to the electronic time resolution $\delta t_e \, (\sim 1\,\mathrm{ns})$ while maintaining a manageable computation time.

To do this, the correlated, non-Poisson distributed photons are artificially diluted with a stream of purely Poisson distributed photons. Contaminating the starlight with purely Poisson light degrades the SNR equivalently to degrading the signal bandwidth, permitting simulations with $\delta t \gg \tau_c$. Without affecting the mean number of photons incident on the telescope, the instantaneous number of photons may be written:
\begin{equation}
n_{T_i}=P(\bar n_{\star_i} (1-\kappa)+\bar \mu_i \kappa + \beta \bar n_{\star_i})
\end{equation}
where the parameter $\kappa$ sets the degree of dilution of the correlated photons without affecting the total photon rate. $\kappa=0$ corresponds to a case in which the correlation is zero, while $\kappa=1$ corresponds to a case of maximal correlation.

We want to obtain the correlation between the signal fluctuations $s_{T_i}$ about the mean. They can be written $s_{T_i}=n_{T_i}-\bar n_{T_i}$, where we use $\bar n_{T_i}=\bar n_{\star_i} (1+\beta)$ so as to ensure $\langle s_{T_i} \rangle = 0$.

The correlation is computed as 
\begin{equation}
g^2=\frac{\langle s_{T_1} \cdot s_{T_2} \rangle}{\bar n_{T_1} \cdot \bar n_{T_2}}. 
\end{equation}
After correction of the mean of $g^2$ for the parameters $\kappa \ \& \ \beta$, (see Appendix \ref{poisstat}), we see that 
\begin{equation}
|\gamma|^2 = \frac{(1+\beta)^2}{\kappa^2} \, g^2
\end{equation}
in the limit where the diameter of the telescopes is small compared to the distance required to begin to resolve surface features of the star. Note that $\kappa=\sqrt{{\tau_c}/{\delta t}}$ is equivalent to the $\sqrt{\Delta f \, T}$ term in Eq.~\ref{SNR} with $\Delta f = 1/{\delta t_e}$.

Photons are recorded using electronic systems with a specific time response. It should be noted that in SII, it is the correlation between the high frequency fluctuations in light intensity recorded by different telescopes that is measured. Such high frequency fluctuations are directly accessible through an AC coupling of the photo-detector. An AC coupled trace can be modeled using any desired single-photon response pulse whose integral is zero. The accumulation of individual photons, each modeled by an AC coupled pulse, directly provides the telescope signal $s_{T_i}$. With a detailed pulse model, the sensitivity of a given system may be evaluated more precisely and the effects of experimental timing inaccuracies may be investigated. In this paper, for the application of our simulations, we restrict ourselves to studying the simplest case of an AC coupled square pulse so that our results can be compared to simple models to validate our approach.

\subsection{Proof of principle of the light model}
\label{proof}
\begin{figure}
  \begin{center}
    \includegraphics[scale=0.35,angle=270]{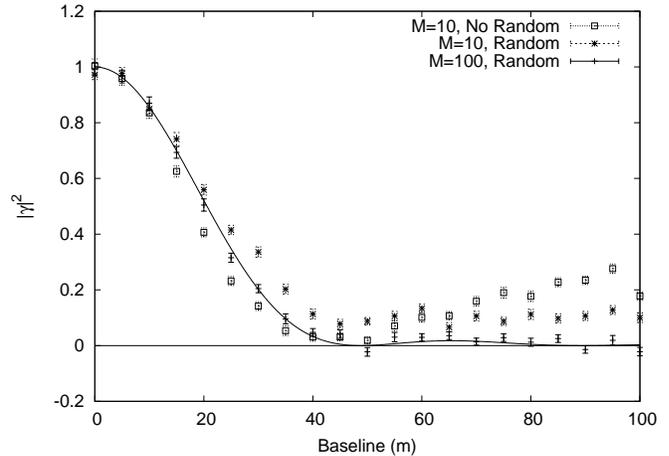}
    \caption{\label{fig:test1}The simulated data for the observation of a uniform disk star, $2\,\rm mas$ in diameter is shown to reproduce an Airy disk profile. See text for details. The data sets shown here are: a star consisting of 10 point sources which are not randomized at each time-step (these points are indicated by $\square$); a star consisting of 10 point sources which are randomized at each time-step (indicated by *); and a star consisting of 100 point sources which are randomized at each time-step (indicated by +). Deviations at large baselines are greatest when the star is simulated with fewer points and when the points of the star are not randomized at each time-step.}
    
  \end{center}
\end{figure}

To test that the simulations yield the correct correlation, we simulate a uniform disk star and calculate $|\gamma|^2$ as a function of the baseline for point-like telescopes, (using a square pulse of width $\delta t_e$), and neglecting detector excess noise and stray light contamination. Figure \ref{fig:test1} shows the simulated data for a uniform disk star, $2\rm \, mas$ in diameter observed with a point-like telescope with a $1{\rm \, m}^2$ flux collection area through a $10\rm\, nm$ optical bandwidth centered on  a wavelength of $400\rm\, nm$. The star, which has a flux $\Phi_\star=10^9\ \mathrm{photons \cdot m^{-2} \cdot s^{-1}}$ through the selected optical bandwidth ($m_V\approx -0.5$), was simulated for a duration of $1 \rm\,ms$. The unrealistically large signal-to-noise ratio is obtained because of the use of a signal bandwidth that is only 10 times smaller than the optical bandwidth. We see that the signal reproduces an Airy disk profile to high precision. However, we found that the fewer the number of point sources making up the star, the greater the power at high frequencies, which causes $|\gamma|^2$ to deviate from the Airy disk profile. Additionally, we found that randomizing the location of each point source within the spatial extension of the star as well as randomizing the frequency emitted by each point source at each time-step is more representative of a realistic star and further reduces deviations from the Airy disk. 

\begin{figure}
  \begin{center}
    \includegraphics[scale=0.35,angle=270]{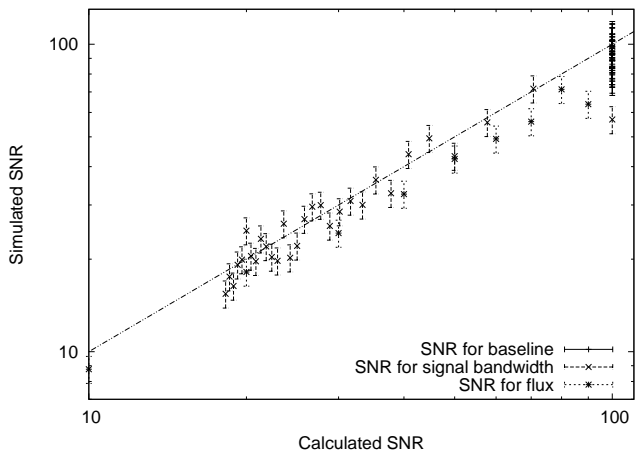}
    \caption{Statistics of simulation results are shown to follow the theoretical signal-to-noise prediction by Hanbury Brown. The + shows data acquired from varying the baseline, the $\times$ shows data acquired from varying the pulse width $\delta t_e$, and the * shows data acquired from varying the flux $\Phi_\star$. The ratio of the simulated to expected SNR for various parameters fall around a 1:1 ratio indicated by the straight line.}
    \label{fig:test2}
  \end{center}
\end{figure}

To further test our approach, we compared the expected SNR calculated from Eq. ~\ref{SNR} to the SNR of simulated correlations. In these experiments, the excess noise and stray light contamination are set to zero. Each numerical experiment was run $50$ times in order to calculate the standard deviation of each set of experimental parameters for a non-resolved star (i.e. $|\gamma|^2=1$). In Figure \ref{fig:test2}, we compare the SNR obtained from simulations to the SNR calculated from Eq. \ref{SNR}. For the signal bandwidth, a square pulse was used so that the width of the pulse is unambiguous. The AC coupling was taken into account by subtracting the average signal rather than including a negative tail for the pulse, so the correlation is obtained between signals $s_{T_i}$ as described in Section \ref{signals}. The width $\delta t_e$ of the pulse was varied from 1$\,\mathrm{ns}$ to 30$\,\mathrm{ns}$ and it is verified that the standard deviation evolved as a square root of the pulse width. Similarly, the flux of the light source was varied from $\Phi_\star=10^8\ \mathrm{photons \cdot m^{-2} \cdot s^{-1}}$ to $\Phi_\star=10^{9}\ \mathrm{photons \cdot m^{-2} \cdot s^{-1}}$. We see that the simulated SNR precisely follows the prescription of Eq. ~\ref{SNR}.  The effect on the SNR of varying the flux is equivalent to varying the light collecting area of the detectors. The effects of increasing the aperture will be further discussed in Section \ref{extended}. These results confirm that our Monte-Carlo approach provides consistent results in simple cases, and so, more subtle instrumentation artifacts can be investigated.

\section{Simulations toward a non-ideal interferometer}
\label{app}

\subsection{Mirror extension}
\label{extended}
When the effect of the telescope mirror extension is important, corrections must be applied to the data before the van Cittert-Zernike Theorem can be used. Simulations can be exploited to establish this correction. The mirror extension effects are significant and corrections must be applied in most cases since telescopes that may be constructed in future arrays such as CTA could be up to $\sim 30\,\mathrm{m}$ in diameter \citep{cta3}, which is comparable to intertelescope distances of current and future arrays.

We have investigated the effect of mirror extension on a $2 \, \mathrm{mas}$ diameter uniform disk star for various telescope diameters which are comparable to the dish sizes considered by CTA. We see that the shape of the ideal curve, which is shown by the Airy disk profile and by the simulated data for the case of a point-like detector (i.e. too small to start resolving the star), is smoothed out as the size of individual detectors increases and begins to resolve the star. 

\begin{figure}
  \begin{center}
    \includegraphics[scale=0.35,angle=270]{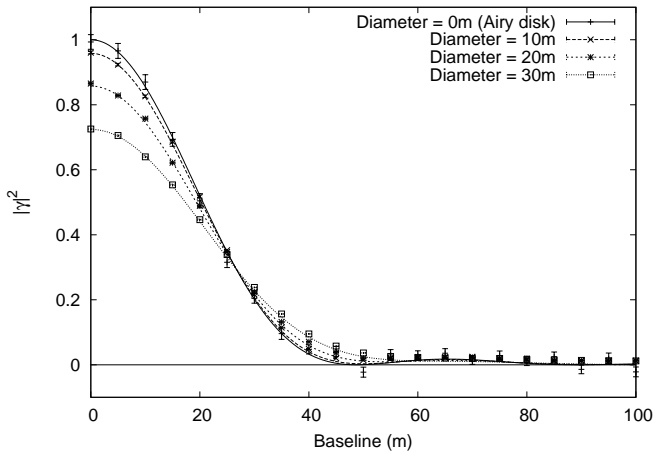}
    \caption{An Airy disk profile is shown by the solid line. The double convolutions of the Airy disk profile with uniform disks of diameters 10 m, 20 m, and 30 m are shown by the dashed lines. The double convoluted profiles are plotted with simulated correlation data for telescopes with mirror diameters of 10 m ($\times$), 20 m (*), and 30 m ($\square$) for comparison. Aside from the telescope extension, the physical parameters used in the simulations are the same as used in Figure \ref{fig:test1}}
    \label{fig:mirrorext_conv}
  \end{center}
\end{figure}

In principle, the effect of large detector sizes on the degree of correlation is equivalent to taking a successive convolution of $|\gamma|^2$ with the shape of the light collecting area of each individual telescope (see Appendix \ref{convolution}) \citep{hbt_book}, which moves $|\gamma|^2$ away from being the squared magnitude of the Fourier transform of the source. Because the simulated star is a uniform disk, $|\gamma|^2$ follows an Airy disk profile independently from the orientation of the direction of the baseline with respect to the star. Therefore, the Airy disk in two dimensions may be obtained by assuming axial symmetry in the correlation data. In order to test the simulation algorithm against the successive convolution model, we simulated a pair of identical telescopes. The double convolutions of the simulated two-dimensional data for two telescopes with uniform disk-shaped light collecting areas of equal diameters, $10 \, \mathrm{m}$, $20 \, \mathrm{m}$ and $30 \, \mathrm{m}$ respectively are shown in Figure \ref{fig:mirrorext_conv}. Within the standard error, each data set agrees well with the prediction.

The effect of mirror extension on image reconstruction capabilities requires a detailed study which cannot be carried out in this paper. It may be possible to develop correction algorithms to apply to the data before analysis to partially alleviate the effect; however, some information is lost in the successive convolutions. Alternatively, when a parameterized image model is available, detailed simulations can be compared to the data for parameter optimization.

\subsection{Excess noise}
In the high-frequency regime of interest here, most of the fluctuations of the telescope signal correspond to the Poisson statistics of the collected photons. To this, we must add the fact that the detector response may fluctuate from photon to photon. This is known as the excess noise.  
For example, in the case of a photo-multiplier, the excess noise results primarily from the fluctuations in the number of electrons ejected at the first dynode. A typical excess noise level for a photo-multiplier is around 30\% of the average single-photon response. The excess noise is uncorrelated between different telescopes but it may have the effect of reducing the SNR achieved in the detection of a correlation between telescopes. Here, we model the excess noise by multiplying each single photon pulse by a Gaussian deviate whose mean is $1$ and whose standard deviation is $\sigma$. However the Gaussian is truncated at zero to avoid multiplication of the pulse by a negative factor. As a consequence, the mean is greater than the most probable amplitude. In order for the mean single photon response amplitude to remain the same, all signals are divided by the mean of the excess noise distribution. The SNR expression in Eq. \ref{SNR} accounts only for the fluctuations associated with photon statistics, so the simulations are used to gain an understanding of the noise introduced by the electronics. Figure \ref{fig:exnoise} shows the SNR obtained from the simulations and from a model developed in Appendix \ref{exn} as a function of the excess noise. The model does not account for the truncation of the distribution of the single photon response pulse at zero. Therefore, as expected, simulated data deviates increasingly from the model as the relative excess noise is increased. The simulated SNR does not degrade as fast as suggested by the model since the distribution of pulse amplitudes is narrower due to the truncation effect. Since the difference in SNR between simulations with $\Phi_\star = 10^6 \ \mathrm{photons \cdot m^{-2} \cdot s^{-1}}$ and $\Phi_\star = 10^7 \ \mathrm{photons \cdot m^{-2} \cdot s^{-1}}$ at zero excess noise is smaller than the statistical error, we increased the number of simulation experiments by a factor of ten while maintaining all the parameters as in Figure \ref{fig:exnoise} in order to establish a clear difference between the sensitivities for these fluxes. We found that from the simulation, the normalized SNR for a flux $\Phi_\star = 10^6 \ \mathrm{photons \cdot m^{-2} \cdot s^{-1}}$ is ${S}/{N}=0.706 \pm 0.007$ compared to a model prediction of ${S}/{N}=0.706$ and for a flux $\Phi_\star = 10^7 \ \mathrm{photons \cdot m^{-2} \cdot s^{-1}}$, the SNR is ${S}/{N}=0.685 \pm 0.007$ compared to a prediction of ${S}/{N}=0.693$. For non-zero excess noise, the SNR departs from the evaluation using Eq. \ref{SNR} as discussed above. This type of study allows for a more accurate and realistic estimate of the sensitivity of an SII experiment. A more realistic model of the excess noise statistics, such as a Polya distribution for photo-multiplier tubes \citep{prescott} can easily be incorporated into the simulation algorithm. Similarly, the noise characteristics of other types of detectors, such as silicon photo-multipliers (SiPM), Geiger-mode avalanche photodiodes, and micro-channel plate (MCP) photo-multipliers \citep{wagner} could be modeled as well. 
\begin{figure}
  \begin{center}
    \includegraphics[scale=0.35,angle=270]{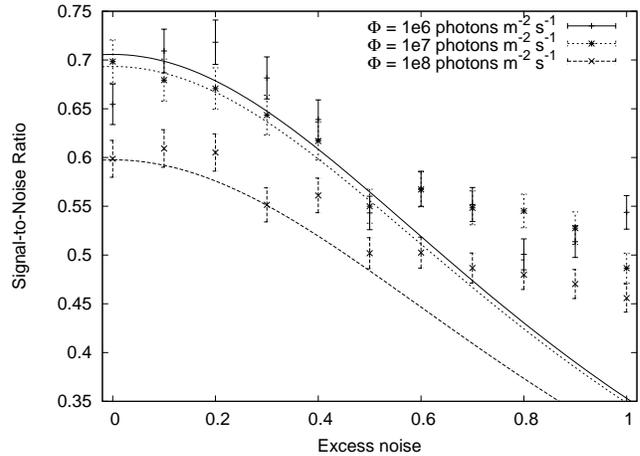}
    \caption{Dependence of SNR on excess noise. Simulations were made using a pair of 1$\, m^2$ telescopes and an observation time of 1ms. For each point, the SNR was obtained from running the simulation 500 times. The + indicates data for fluxes of $\Phi_\star = 10^6 \, \mathrm{photons \cdot m^{-2} \cdot s^{-1}}$ (approximately corresponding to a visual magnitude $m_V\approx 7$ through a $10{\rm nm}$ optical bandwidth), * indicates data for fluxes of $\Phi_\star = 10^7 \, \mathrm{photons \cdot m^{-2} \cdot s^{-1}}$ ($m_V\approx 4.5$), and $\times$ indicates data for fluxes of $\Phi_\star = 10^8 \, \mathrm{photons \cdot m^{-2} \cdot s^{-1}}$ ($m_V\approx 2$) and the corresponding solid and dashed lines indicate the respective models. The model of the SNR fits the simulated data well for lower values of excess noise.}
    \label{fig:exnoise}
  \end{center}
\end{figure}

\subsection{Night sky contamination}
\begin{figure}
  \begin{center}
    \includegraphics[scale=0.35,angle=270]{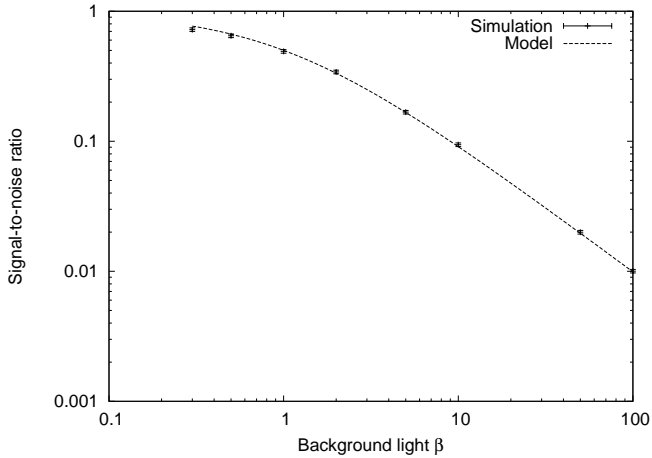}
    \caption{Dependence of SNR on the amount of background light $\beta$ in terms of the source flux $\Phi_\star=10^8 \, \mathrm{photons \cdot m^{-2} \cdot s^{-1}}$ from an unresolved source observed  for a duration of $T=0.1 \, \mathrm{\mu s}$ with $1{\rm \, m}^2$ light collecting area telescopes. For each point the SNR was obtained from running the simulation 500 times.} 
    \label{fig:bglight}
  \end{center}
\end{figure}

The simulations can also be used to investigate the effects of stray light contamination. When very faint stars are to be observed, the observation time must be long enough to obtain a sufficient SNR for the measurement. In these cases, contamination from the night sky background, especially during bright moonlit nights, cannot be ignored. When the light received is dominated by the night sky background contamination, increasing the observation time is no longer beneficial. An arbitrary amount of stray light can be included in the simulation as a stream of additional photons with purely Poisson statistics. The expected dependence of the SNR on the amount of stray light is derived in Appendix \ref{poisstat}. To test this, we simulated a star of brightness $\Phi_\star=10^8 \, \mathrm{photons \cdot m^{-2} \cdot s^{-1}}$ and integrated over $T=0.1 \, \mathrm{\mu s}$. The dependence of the SNR as a function of the stray light, in terms of the brightness of the source, is shown in Figure \ref{fig:bglight}. We observe a good match between the simulated data and our model.

\begin{figure}
  \begin{center}
    \includegraphics[scale=0.35,angle=270]{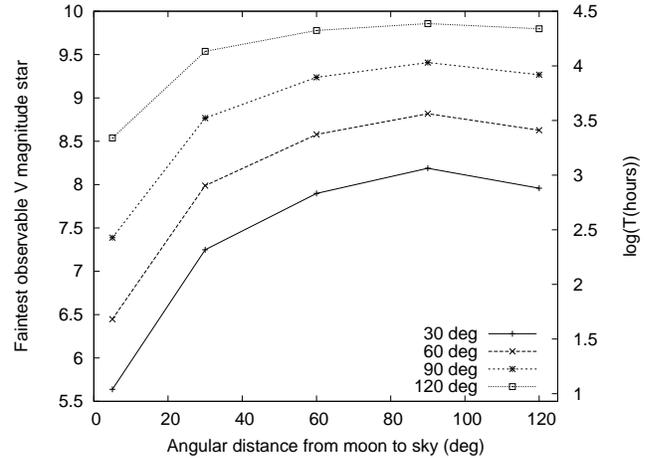}
    \caption{Faintest stars that can be observed using SII with telescopes with a PSF of $0.06^{\circ}$ FWHM under various moonlight conditions. The horizontal axis shows the angular distance of the star from the moon, the vertical axis shows the faintest visual magnitude star that can be observed, and the various curves correspond to different phases of the moon, where + indicates a moon phase of $30 \, \mathrm{deg}$, $\times$ indicates a moon phase of $60 \, \mathrm{deg}$, * indicates a moon phase of $90 \, \mathrm{deg}$, and $\square$ indicates a moon phase of $120 \, \mathrm{deg}$. The right vertical axis shows the integration time required to achieve a $SNR=5$ for a star with spectral density $\eta_V=(5 \cdot 10^{-5} \, \mathrm{photons \cdot m^{-2} \cdot s^{-1} \cdot Hz^{-1}) \cdot 2.5^{-V}}$ at $|\gamma|^2=0.5$ using a telescope of area $A=100\, \mathrm{m^2}$, a photodetector quantum efficiency of $\alpha=0.25$, and a signal bandwidth of $\Delta f = 100 \, \mathrm{MHz}$.}
    \label{fig:moonlight}
  \end{center}
\end{figure}

Stray light contamination ultimately limits the faintness of stars that can be observed. Using the brightness of the night sky for various phases and distances from the moon from \cite{krisciunas}, Figure \ref{fig:moonlight} shows the magnitude of the stars whose brightness equals the night sky luminosity integrated within the PSF. This situation corresponds to $\beta=1$ which results in a degradation of the SNR by a factor of $2$ compared to $\beta=0$ (i.e. no background light). We used an optical PSF full width at half maximum (FWHM) of $0.06^{\circ}$, corresponding to the VERITAS telescope array \citep{holder}. Another limitation for the observability of a star is the practicality of the required integration time. For this reason, in Figure \ref{fig:moonlight} we include the integration time required to achieve a $SNR=5$ for a star with spectral density $\eta_V=(5 \cdot 10^{-5} \, \mathrm{photons \cdot m^{-2} \cdot s^{-1} \cdot Hz^{-1}) \cdot 2.5^{-V}}$ at $|\gamma|^2=0.5$ using a telescope of area $A=100\, \mathrm{m^2}$, a photodetector quantum efficiency of $\alpha=0.25$, and a signal bandwidth of $\Delta f = 100 \, \mathrm{MHz}$. This does not take into account the fact that in a telescope array, the redundancy of the baselines can be used to improve the sensitivity.

\begin{figure}
  \begin{center}
    \includegraphics[scale=0.35,angle=270]{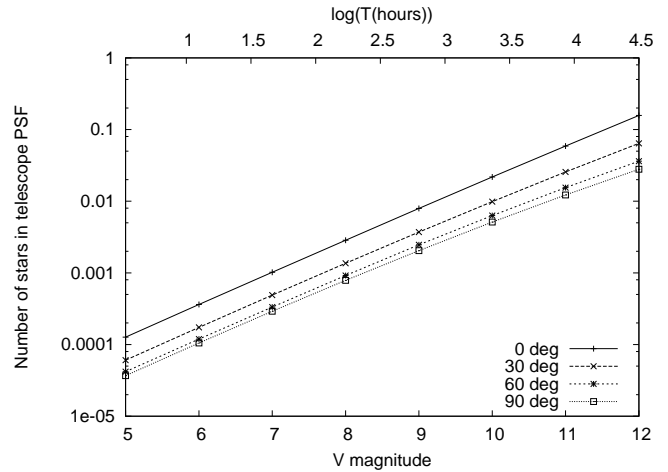}
    \caption{The vertical axis shows the average number of stars in the field of view of a telescope with a PSF FWHM of $0.06^{\circ}$ as a function of the apparent visual magnitude $V$. The various curves correspond to the angular distances from the galactic plane, where + corresponds to an angular distance of $0 \, \mathrm{deg}$, $\times$ corresponds to an angular distance of $30 \, \mathrm{deg}$, * corresponds to an angular distance of $60 \, \mathrm{deg}$, and $\square$ corresponds to an angular distance of $90 \, \mathrm{deg}$. The upper horizontal axis shows the integration time required to achieve a $SNR=5$ for a star with spectral density $\eta_V=(5 \cdot 10^{-5} \, \mathrm{photons \cdot m^{-2} \cdot s^{-1} \cdot Hz^{-1}) \cdot 2.5^{-V}}$ at $|\gamma|^2=0.5$ using a telescope of area $A=100\, \mathrm{m^2}$, a photodetector quantum efficiency of $\eta=0.25$, and a signal bandwidth of $\Delta f = 100 \, \mathrm{MHz}$.}
    \label{fig:stellardens}
  \end{center}
\end{figure}

In addition to limitations imposed by the moon, SII measurements may also be affected by other stars in the field of view of the telescope, especially when observing very faint stars. If there is a star which is as bright or brighter than the star of interest in the PSF of the telescope then it is no longer possible to measure the star of interest. As the brightness of the star decreases then the average number of stars within the PSF of the telescope increases, as shown in Figure \ref{fig:stellardens}. However, for stars brighter than magnitude 12, given the low density of bright stars, other stars within the telescope PSF are unlikely to interfere with measurements. Stars of magnitude 12 or fainter will likely remain out of reach of SII because of the required integration time.

\section{Conclusion}
\label{conclu}
Previous studies have not accounted for many realistic effects associated with SII measurements. Here, we have demonstrated that our simulations correctly reproduce our simple models predicting how $|\gamma|^2$ and the measurement sensitivities evolve with various parameters, and we have briefly discussed instrumental properties such as telescope mirror extension, signal bandwidth limitation and electronic pulse shape, excess noise, and stray light contamination. We have verified that the signal measured with spatially extended detectors is the successive convolutions of the normalized Fourier Transform of the source with the shape of each detector area. We also verified that the SNR degrades as the excess noise increases according to a simple model developed in Appendix \ref{exn}. Additionally, we find that the sensitivity degradation departs from our model for large values of excess noise, also as expected. Finally, we see that when the contamination of starlight by the night sky background becomes important, the SNR degrades so that increasing the observation time no longer offers benefits to the measurements. Again we find that the SNR degrades according to our simple model developed in Appendix \ref{poisstat}. These tests and comparisons of simulation results and simple models were made while isolating specific instrumental aspects one at a time. The good agreements obtained in all cases gives us confidence that the simulations may be used to characterize the performances of realistic instruments via their detailed modeling. This is important to gain an understanding of the sensitivity of existing and planned instruments and also this may be used to develop data correction algorithms so as to alleviate instrumental effects impacting the signals. 

Many other effects that will be encountered during real SII measurements can also be investigated with the simulations. For example, the simulations can be used to investigate the effects of inaccuracies in the time alignment of the signals. Furthermore, ACT optics are generally of Davies-Cotton design \citep{davis-cotton1957}, which is not isochronous. In previous studies, this was approximately accounted for as a signal bandwidth limitation while we could now simulate non-isochronous effects in detail.

A possible implementation approach to SII consists in digitizing the individual telescope signals. The required fast digitization rate, exceeding 100 mega-samples per second, implies that huge volumes of data must be handled. However, this provides a lot of flexibility as the correlation is obtained off-line at the time of data analysis. The simulation could also incorporate a model of the digitizer.

The simulation algorithm presented in this paper can in fact be combined with any details of the envisioned instrumentation to develop and optimize data analysis and characterize the performance of any stellar intensity interferometer.

\section{Acknowledgements}
The authors are thankful to Michael Daniel for his time and the helpful suggestions he made to improve the clarity of this paper.

\appendix

\section{Partially correlated Poisson Statistics}
\label{poisstat}
The total number of photons (i.e. from the star and from the background light) incident on a telescope $i$ in a time $\delta t$  is given by
\begin{eqnarray}
n_{T_i} &=& P(\bar n_{\star_i} (1-\kappa) + \bar \mu_i \kappa + \bar n_{BG_i}) \nonumber \\
&=& P(\bar n_{\star_i} (1-\kappa) + \bar \mu_i \kappa + \beta \bar n_{\star_i}) \label{eq:nTi}
\end{eqnarray}
where $P(m)$ represents a Poisson distribution of mean and variance $m$, $\bar n_{\star_i}$ is the mean number of photons emitted by the star and collected by telescope $i$, and $n_{BG_i} := \beta \bar n_{\star_i}$ with $\beta$ being a dimensionless factor which provides a measure of the amount of background light in terms of the amount of light from the star.
 
The mean number of photons incident on a telescope is the sum of the photons from the star and from the night sky:
\begin{equation}
\bar n_{T_i} = \bar n_{\star_i}\,(1+\beta) \label{eq:nTbar}
\end{equation}
where $\bar n_{\star i}=\Phi_\star \, A \, \delta t$.

From equations \ref{eq:nTi} and \ref{eq:nTbar} the signal from telescope $i$ is
\begin{eqnarray}
s_{T_i} = n_{T_i} - \bar n_{T_i} = P(\bar n_{\star_i} (1-\kappa) + \bar \mu_i \kappa + \beta \bar n_{\star_i}) - \bar n_{\star_i}\,(1+\beta) \nonumber
\end{eqnarray}
so that $\langle s_{T_i} \rangle = 0 $.

The correlation $g^2$ between the two telescopes signals is:
\begin{equation}
\label{eq:simg}
g^2 = \frac{\langle s_{T_1} \cdot s_{T_2}  \rangle}{\bar n_{\star_1} \bar n_{\star_2} (1+\beta)^2}.
\end{equation}
Considering just the numerator:

\begin{eqnarray}
\langle s_{T_1} \cdot s_{T_2}  \rangle &=& \langle (n_{T_1}-\bar n_{T_1}) \cdot (n_{T_2} - \bar n_{T_2}) \rangle \nonumber \\
&=& \langle (P(\bar n_{\star_1} (1-\kappa) + \bar \mu_1 \kappa + \beta \bar n_{\star_1}) - \bar n_{\star_1}\,(1+\beta)) \nonumber \\
&\ & \cdot (P(\bar n_{\star_2} (1-\kappa) + \bar \mu_2 \kappa + \beta \bar n_{\star_2}) - \bar n_{\star_2} \,(1+\beta))  \rangle \nonumber
\end{eqnarray}

Using Raikov's Theorem, which states that for a random variable $x_i$, if
\begin{equation}
x_i = P(m_i)\ {\rm then}\ \sum x_i = P\left( \sum m_i \right),
\end{equation}
then $\langle s_{T_1} \cdot s_{T_2}  \rangle$ can be rewritten:
\begin{eqnarray}
\langle s_{T_1} \cdot s_{T_2} \rangle &=& \langle (P(\bar n_{\star_1} (1-\kappa)) + P(\bar \mu_1 \kappa) + P(\beta \bar n_{\star_1}) \nonumber \\
&\ & - \bar n_{\star_1} \, (1+\beta)) \nonumber \\
&\ & \cdot (P(\bar n_{\star_2} (1-\kappa)) + P(\bar \mu_2 \kappa) + P(\beta \bar n_{\star_2}) \nonumber \\
&\ &  \bar n_{\star_2} \, (1+\beta)) \rangle \nonumber 
\end{eqnarray}
Expanding and proceeding with the calculation by considering each term separately, it is found that:
\begin{eqnarray}
\langle s_{T_1} \cdot s_{T_2} \rangle &=& \kappa^2 \langle \bar \mu_1 \cdot \bar \mu_2 \rangle -{\bar n_{\star_1}}{\bar n_{\star_2}}\,\kappa^2 \nonumber \\
&=& \kappa^2 \langle (\bar \mu_1 - \bar n_{\star_1}) \cdot (\bar \mu_2 - \bar n_{\star_2}) \rangle \label{eq:s1s2}
\end{eqnarray}
recalling the fact that $\langle \bar \mu_i \rangle=\bar n_{\star_i}$.

The simulation correlation is found by substituting equation \ref{eq:s1s2} into equation \ref{eq:simg}:
\begin{equation}
g^2 = \frac{\langle s_{T_1} \cdot s_{T_2} \rangle}{\bar n_{\star_1} \bar n_{\star_2} (1 + \beta)^2} = \frac{\kappa^2 \langle (\bar \mu_1 - \bar n_{\star_1}) \cdot (\bar \mu_2 - \bar n_{\star_2}) \rangle}{\bar n_{\star_1} \bar n_{\star_2} (1 + \beta)^2} \nonumber
\end{equation}

Using the definition of the squared degree of coherence,
\begin{equation}
|\gamma|^2 = \frac{\langle (\bar \mu_1 - \bar n_{\star_1}) \cdot (\bar \mu_2 - \bar n_{\star_2}) \rangle}{\bar n_{\star_1} \bar n_{\star_2}}
\end{equation}
then
\begin{equation}
g^2 = \frac{\kappa^2 |\gamma|^2}{(1+\beta)^2} \nonumber
\end{equation}

and solving for the quantity of interest, $|\gamma|^2$:
\begin{equation}
|\gamma|^2 = \frac{(1+\beta)^2}{\kappa^2} \, g^2
\end{equation}

where $\kappa := \sqrt{\frac{\tau_c}{\delta t}}$.

Since only the starlight may present correlation, the signal-to-noise expression used in Figure \ref{fig:bglight} which accounts for stray light contamination is derived from Equation \ref{eq:sgnl} with $\eta=\eta_{\star}$ where $\eta_{\star}$ is the spectral density of the star, and Equation \ref{eq:noise} with $\eta=\eta_{\star}(1+\beta)$ so that:
\begin{equation}
\frac{S}{N}=A\,\alpha \left( \frac{\eta_{\star}}{1+\beta} \right)  |\gamma|^2 \sqrt{\Delta\,f\,T/2}.
\end{equation}

\section{Effect of detector area on the complex degree of coherence}
\label{convolution}
The effect of large telescope area may be quite significant in practice, so it is of interest to be able to predict how it affects and degrades the signal. To do this, first we set the origin of the coordinate system at the light source. Points on the source are labeled by positions $\vec{x}'$ with respect to the origin. The vector radii to the same points from a far away observer at position $\vec{x}$ are denoted:
\begin{equation}
 \vec{r}=\vec{x}-\vec{x}'.
\end{equation}
For a point-like detector, the observed amplitude at $\vec{x}$ is
\begin{equation}
  E(\vec{x},t)=\int \frac{A(\vec{x}',t)}{|\vec{r}|}e^{i(kr-\omega t+\phi(\vec{r},t))}d^2x',
\end{equation}
where $\phi(\vec{r},t)$ is a random phase caused by atmospheric turbulence among other factors. The following approximation can be made:
\begin{equation}
  |\vec{r}|\approx |\vec{x}|-\vec{x}'\cdot \frac{\vec{x}}{|\vec{x}|},
\end{equation}
so that 
\begin{equation}
  E(\vec{x},t)=\frac{e^{ikx}}{|\vec{x}|}\int A(\vec{x}',t)\,e^{\left\{i\left(k\vec{x}'\cdot \frac{\vec{x}}{|\vec{x}|}-\omega t+\phi(\vec{r},t) \right)\right\}}d^2x'.
\end{equation}
If the detector has a finite area, then the amplitude at position $\vec{x}$ is a superposition of amplitudes at positions $\vec{x}-\vec{x}_d$, where $\vec{x}_{d}$ are points in the detector with respect to position $\vec{x}$. The random phase can be expressed as a function of detector coordinates as $\phi(\vec{x}_d,t)$.  Now the superposition of amplitudes is expressed as a convolution with the detector area, i.e.
\begin{eqnarray}
  E(\vec{x},t)\approx \frac{e^{ikx}}{|\vec{x}|}\int A(\vec{x}',t)\,e^{\left\{i\left(k\vec{x}'\cdot \frac{\vec{x}-\vec{x}_d}{|\vec{x}|}-\omega t+\phi(\vec{x}_d,t)\right)\right\}}d^2x'd^2x_d. \nonumber
\end{eqnarray}
To calculate the time averaged correlation between detectors $i$ and $j$, denoted as $\left\langle E(\vec{x}_i) E^*(\vec{x}_j)\right\rangle$, note that 
\begin{eqnarray}
  \left\langle A(\vec{x}',t)A(\vec{x}'',t)e^{i(\phi(\vec{x}_{di},t)-\phi(\vec{x}_{dj},t))}\right\rangle \nonumber \\ 
=I(\vec{x}')\delta(\vec{x}'-\vec{x}'')\delta(\vec{x}_{di}-\vec{x}_{dj}), \nonumber
\end{eqnarray}
where $I(\vec{x}')$ is the light intensity at point $\vec{x}'$. This is because separate points on the source are not correlated over large distances. The phase is also not correlated between separate points, that is, $\phi(\vec{x}_{di},t)-\phi(\vec{x}_{dj},t)$ will only be zero when ${x}_{di}={x}_{dj}$; otherwise it will have a time variation which results in $  \left\langle e^{i(\phi(\vec{x}_{di},t)-\phi(\vec{x}_{dj},t))} \right\rangle=0$ when ${x}_{di}\neq{x}_{dj}$.\\
Now defining $\vec{z}_i\equiv \vec{x}_i-\vec{x}_{di}$, the time averaged correlation is
\begin{eqnarray}
  \left\langle E(\vec{x}_i) E^*(\vec{x}_j)\right\rangle =C \int I(\vec{x}')\,e^{\left\{ik\left(\vec{x}'\cdot \frac{\vec{z}_i}{|\vec{x}_i|}-\vec{x}'\cdot \frac{\vec{z}_j}{|\vec{x}_j|}\right)\right\}} d^2x' d^2x_{dj}. \nonumber
\end{eqnarray}
where $C$ is a constant. When $|\vec{x}'|\ll |\vec{x}_j|$, then the angle $\vec{\theta}$ can be defined as 
\begin{equation}
  \vec{\theta}\equiv \frac{\vec{x}'}{|\vec{z}_j|},
\end{equation}
The correlation can now be expressed as:
\begin{eqnarray}
\left\langle E(\vec{x}_i) E^*(\vec{x}_j)\right\rangle&=& \int I(\vec{\theta})\,e^{-ik\vec{\theta}\cdot(\vec{z}_i-\vec{z}_j)}\, d^2\theta \, d^2x_{dj} \\
&=&\int \tilde{I}(\vec{z}_i-\vec{z}_j)\, d^2x_{dj},
\end{eqnarray}
where $\tilde{I}(\vec{z}_i-\vec{z}_j)$ is the Fourier transform of the radiance distribution of the star, which goes from angular space to detector separation space. Now the quantity measured in intensity interferometry is 
\begin{eqnarray}
  |\gamma(\vec{x_i},\vec{x_j})|^2 &=& \frac{|\left\langle E(\vec{x}_i) E^*(\vec{x}_j)\right\rangle|^2}{\sqrt{|E(\vec{x}_i)|^2 |E(\vec{x}_j)}|^2}\\
  &=& \frac{1}{I(\vec{x}_i)I(\vec{x}_j)}\int |\tilde{I}(\vec{z}_i-\vec{z}_j)|^2\, d^2x_{di}d^2x_{dj}. \nonumber
\end{eqnarray}
Therefore, the effect of having finite sized telescopes is to replace the magnitude of mutual degree of coherence $|\gamma|^2$ by its successive convolutions with each of the telescope light collection area shapes.

\section{Excess Noise}
\label{exn}
Ignoring the effects of stray light, the number of photons during one time-step $\delta t$ in channel $i$ is
\begin{equation}
n_i=P(\bar n_i (1-\kappa) + \bar \mu_i \kappa).
\end{equation}
where $P(m)$ represents a Poisson distribution of mean and variance $m$.

The correlation is a result of the non-Poisson term, $\bar \mu_i$. The number of photons incident on channel $i$ can also be written as the sum of a Poisson term and a Binomial term as follows:
\begin{equation}
n_i=P(\bar n_i (1-\chi)) + B(\chi)n_j
\end{equation}
where $\chi$ is the probability that when a photon arrives in channel $i$ there is a correlated photon that also arrives in channel $j$, $\langle n_i \rangle = \bar n_i$ and for simplicity we set $n_i \in \{0,1\}$, restricting ourselves to a situation in which $\langle n_i \rangle \ll 1$.

The excess noise is introduced as a Gaussian variation in the single photon response pulse. Our model simply multiplies the amplitude by a random Gauss variable of mean 1 and standard deviation $\sigma$:
\begin{equation}
s_i=\sum^{k=n_i}_{k=1}G(1,\sigma) - \bar n_i
\end{equation}
where the signal $s_i$ is made to be AC coupled by subtracting the mean number of photons $\bar n$ from the sum of individual photon signals (each of mean $1.0$) so that the mean of the signal is $0$. 

Since the number of photons incident on a telescope per time-step is considered to be small ($n_i \in \{0,1\}$), the signal can be rewritten as
\begin{equation}
s_i \simeq n_i \, G(1,\sigma) - \bar n.
\end{equation}

The quantity of interest is the effect that the Gaussian factor has on the sensitivity of the measurement, so standard deviation of the following term will be calculated:
\begin{equation}
|\gamma|^2=\frac{\langle s_1 \cdot s_2 \rangle}{\langle n_1 \rangle \, \langle n_2 \rangle}
\end{equation}
with
\begin{equation}
Stdev(|\gamma|^2) = \frac{Stdev(\langle s_1 \cdot s_2 \rangle)}{\bar n_1 \, \bar n_2}.
\end{equation}
To lighten notation we assume that the two telescopes are identical and receive the same amount of light from the star (i.e. $\bar n_1 = \bar n_2 = \bar n$).

Developing the product of the signals,
\begin{eqnarray}
s_1 \cdot s_2 &=& (n_1 \, G(1,\sigma) - \bar n) \cdot (n_2 \, G(1,\sigma) - \bar n) \\ &=& (P(\bar n (1-\chi)) \, G(1,\sigma) + B(\chi) \, n_2 \, G(1,\sigma) - \bar n) \nonumber \\ & \  & \cdot (P(\bar n (1-\chi)) \, G(1,\sigma) + B(\chi) \, n_1 \, G(1,\sigma) - \bar n) \nonumber.
\end{eqnarray}
The effect of the excess noise on a correlated signal should be equivalent to the effect on an uncorrelated signal, i.e. $\chi = 0$. Applying that, the correlation can be rewritten as:
\begin{eqnarray}
s_1 \cdot s_2 &=& (P(\bar n) \, G(1,\sigma) - \bar n) \cdot (P(\bar n) \, G(1,\sigma) - \bar n) \\ &=& P(\bar n) \, G(1,\sigma) \, P(\bar n) \, G(1,\sigma)  + {\bar n}^2 \nonumber \\ & \  & - P(\bar n) \, G(1,\sigma) \, \bar n  - P(\bar n) \, G(1,\sigma) \, \bar n.  \nonumber
\end{eqnarray}

Note: Suppose $X$ and $Y$ are random independent variables then \citep{ahn}
\begin{eqnarray}
\label{eq:variance}
Var(XY) &=& E^2(X) \, Var(Y) + E^2(Y) \, Var(X) \nonumber \\
&\ &+ Var(X) \, Var(Y)
\nonumber
\end{eqnarray}
The variance of each term can be calculated individually, resulting in
\begin{eqnarray}
Var(s_1 \cdot s_2) &=& \sigma^4({\bar n}^4 + 2 {\bar n}^3 + {\bar n}^2) + \sigma^2 (2 {\bar n}^4 + 4 {\bar n}^3 + 2 {\bar n}^2) \nonumber \\
&\ & + (2 {\bar n}^3 + {\bar n}^2) + 2 (\sigma^2 ({\bar n}^4 + {\bar n}^3) + {\bar n}^3) \nonumber \\
&=& \sigma^4 ({\bar n}^4 + 2 {\bar n}^3 + {\bar n}^2) + \sigma^2 (4 {\bar n}^4 + 6 {\bar n}^3 + 2 {\bar n} ^2) \nonumber \\
&\ & + (4 {\bar n}^3 + {\bar n}^2) \nonumber
\end{eqnarray}
Then,
\begin{eqnarray}
Var(|\gamma|^2) &=& \frac{Var(s_1 \cdot s_2)}{{\bar n_1}^2 \, {\bar n_2}^2} 
\end{eqnarray}
And
\begin{eqnarray}
\label{eq:stdev}
Stdev(|\gamma|^2) &=& \frac{1}{{\bar n}^2} \, [\sigma^4 ({\bar n}^4 + 2 {\bar n}^3 + {\bar n}^2) \\
&\ & + \sigma^2 (4 {\bar n}^4 + 6 {\bar n}^3 + 2 {\bar n} ^2) + (4 {\bar n}^3 + {\bar n}^2)]^{\frac{1}{2}} \nonumber
\end{eqnarray}

The standard deviation in equation \ref{eq:stdev} is the standard deviation of just one time-step, while the term of interest is the standard deviation of the entire measurement, or the standard deviation of the mean:

\begin{eqnarray}
Stdev(|\gamma|^2) &=& \frac{1}{\sqrt{N}{\bar n}^2} \, [\sigma^4 ({\bar n}^4 + 2 {\bar n}^3 + {\bar n}^2) \\
&\ & + \sigma^2 (4 {\bar n}^4 + 6 {\bar n}^3 + 2 {\bar n} ^2) + (4 {\bar n}^3 + {\bar n}^2)]^{\frac{1}{2}} \nonumber
\end{eqnarray}
where $N=\frac{T}{\delta t}$ is the number of time-steps taken in the simulation.

Note that in the calculation described above, the Gaussian variable is truncated at zero to avoid multiplying the pulse by a negative number. This becomes more significant for large values of excess noise $\sigma$ while it is negligible when $\sigma \ll 1$.

\bsp

\label{lastpage}

\end{document}